\documentclass[twocolumn,preprintnumbers,showpacs,prl,floatfix,superscriptaddress,amsmath,amssymb]{revtex4}
\usepackage{epsfig}
\usepackage{subfigure}
\usepackage{amsmath}
\usepackage{color}
\usepackage{amssymb}
\usepackage{verbatim}
\usepackage{setspace}
\usepackage{graphicx}% Include figure files
\usepackage{dcolumn}% Align table columns on decimal point
\usepackage{bm}% bold math
\usepackage{times}
\input epsf

\graphicspath{{figures/}}

\begin{document}

\author{Per Sebastian Skardal}
\email{skardal@colorado.edu} 
\affiliation{Department of Applied Mathematics, University of Colorado at Boulder, Colorado 80309, USA}

\author{Alain Karma}
\affiliation{Physics Department and Center for Interdisciplinary Research on Complex Systems, Northeastern University, Boston, MA 02115, USA}

\author{Juan G. Restrepo} 
\affiliation{Department of Applied Mathematics, University of Colorado at Boulder, Colorado 80309, USA}

\title{Unidirectional Pinning and Hysteresis of Spatially Discordant Alternans in Cardiac Tissue}

%\date{\today}

\begin{abstract}
Spatially discordant alternans is a widely observed pattern of voltage and calcium signals in cardiac tissue that can precipitate lethal cardiac arrhythmia. Using spatially coupled iterative maps of the beat-to-beat dynamics, we explore this pattern's dynamics in the regime of a calcium-dominated period-doubling instability at the single cell level. We find a novel nonlinear bifurcation associated with the formation of a discontinuous jump in the amplitude of calcium alternans at nodes separating discordant regions. We show that this jump unidirectionally pins nodes by preventing their motion away from the pacing site following a pacing rate decrease, but permitting motion towards this site following a rate increase. This unidirectional pinning leads to strongly history-dependent node motion that is strongly arrhythmogenic.
\end{abstract}


\maketitle

The study of period-two dynamics in cardiac tissue has become an important topic of research in the physics~\cite{KarmaGilmour} and biomedical communities~\cite{saga}. The term alternans describes beat-to-beat alternations of both action potential duration (APD) and peak intracellular calcium concentration ($\rm Ca_i^{peak}$). Heart cells generically exhibit alternans when they are paced rapidly or in pathological conditions.  Interest in alternans during the last decade has stemmed from the discovery that APD alternations can become ``spatially discordant'' in tissue~\cite{Pastore}, meaning that APD alternates with opposite phases in different regions~\cite{Hayashi,ZivMironov}. Spatially discordant alternans (SDA) dynamically creates spatiotemporal dispersion of the refractory period during which cells are not excitable, thereby promoting wave blocks and the onset of lethal cardiac arrhythmias~\cite{saga}.   

To date, our theoretical understanding of SDA is well developed for the case where APD alternans results from an instability of membrane voltage ($V_m$) dynamics at the single cell level, which originates from the restitution relation between APD and the preceding diastolic interval (DI) between two action potentials. Numerical simulations~\cite{disc} have shown that ``nodes'', which are line defects with period-1 dynamics separating discordant regions of period-2 oscillations of opposite phases,  can form spontaneously in paced homogeneous tissue due to conduction velocity (CV) restitution, the relationship between action potential propagation speed $cv$ and DI. In addition, node formation has been understood theoretically in an amplitude equation framework~\cite{EK,Dai} to result from a pattern forming linear instability that amplifies spatially periodic stationary or traveling modulations of alternans amplitude. 

Despite this progress, our theoretical understanding of SDA remains incomplete. Both experiments~\cite{Chudin} and ionic model simulations~\cite{num} have shown that $\rm Ca_i^{peak}$ can alternate even when $V_m$ is forced to be periodic with a clamped action potential waveform, demonstrating that alternans can also result from an instability of intracellular calcium dynamics. Although alternans are presently believed to be predominantly $\rm Ca_i$-driven in many instances, our understanding of nodes in this important case remains limited. Numerical simulations have shown a qualitatively similar role of CV-restitution in SDA formation for $V_m$- and $\rm Ca_i$-driven alternans~\cite{Hayashi,Sato06}, but more complex behaviors for the latter case depending on the strength of $\rm Ca_i$-driven instability~\cite{Sato07} and the nature of $\rm Ca_i$-$V_m$ coupling~\cite{Sato06,Zhao}. However, a theoretical framework to interpret both computational and experimental observations has remained lacking. 

In this Letter, we extend the theoretical framework of~\cite{EK} to uncover novel aspects of SDA formation for $\rm Ca_i$-dominated instability and validate our theoretical predictions with detailed ionic model simulations. A major finding is that node motion can be pinned in one direction owing to the formation of a discontinuous jump in calcium alternans amplitude at a node where only $V_m$ exhibits period-1 dynamics. This jump leads to strongly history-dependent SDA evolution and also alters fundamentally the spacing between nodes. We summarize here our main results and additional details of both theory and simulations will be described elsewhere.

We start our analysis from the system of spatially coupled maps of the general form 
\begin{align}
C_{n+1}(x) &= f_c[C_n(x), D_n(x)],\label{gmap1}\\
A_{n+1}(x) &= \int_0^L G(x,x')f_a[C_{n+1}(x'), D_n(x')]dx'\label{gmap2},
\end{align}
where $A_n(x)$, $D_n(x)$, and $C_n(x)$ denote the APD, DI, and $\rm Ca_i^{peak}$, respectively, at beat $n$ and position $x$, and $G(x,x')$ captures the cumulative effect of electrotonic ($V_m$-diffusive) coupling during one beat.  For a cable of length $L$ paced at $x=0$ with no flux boundary conditions on $V_m$ at both ends,  
$G(x,x')=G(x-x')+G(x+x')+G(2L-x-x')$, with $G(x)=H_\xi(x)\left[1+\frac{wx}{\xi^2}\left(1-\frac{x^2}{\xi^2}\right)\right]$ where $H_\xi$ is Gaussian with standard deviation $\xi$ 
(see Appendix B of ~\cite{EK})
%\begin{equation}\label{eqG}
%G(x)=\frac{1}{\sqrt{2\pi \xi^2}}e^{-\frac{x^2}{2\xi^2}}\left[1+\frac{wx}{\xi^2}\left(1-\frac{x^2}{\xi^2}\right)\right],
%\end{equation}
and $\xi=\sqrt{2D_V \text{APD}^*}$ and $w=2D_V/cv^*$ are two intrinsic lengthscales expressed in terms of the APD and CV at the alternans bifurcation (APD$^*$ and $cv^*$, respectively), and $D_V$ is the diffusion constant of $V_m$ in the standard cable equation $\dot V_m=D_V\partial_x^2V_m-I_{\rm ion}$. Furthermore, CV-restitution causes the activation interval $T_n(x)\equiv A_n(x)+D_n(x)$ to vary from beat to beat along the cable as \cite{Courtemanche93,EK}
\begin{equation}
T_n(x)=\tau+ \int_0^x\frac{dx'}{cv(D_n(x'))} - \int_0^x\frac{dx'}{cv(D_{n-1}(x'))},\label{period}
\end{equation}
where $\tau$ is the imposed period at the paced end ($x=0$). To complete the model, we need to specify the forms of $f_a$ and $f_c$. 
Since we are interested in understanding the generic behavior of alternans, we choose simple phenomenological forms of those maps defined implicitly by 
\begin{eqnarray}
f_c/C^*&=&1-rc_n+c_n^3+\alpha d_n,\label{map1}\\
f_a/A^*&=&1+\beta d_n+\gamma c_{n+1},\label{map2}
\end{eqnarray}
where $c_n\equiv (C_n-C^*)/C^*$, $d_n=(D_n-D^*)/A^*$, and we also define $a_n=(A_n-A^*)/A^*$. With this choice $C_n=C^*$, $A_n=A^*$, and $D_n=D^*$ are trivial fixed points corresponding to $c_n=a_n=d_n=0$. Moreover, $c_n$, $a_n$, and $d_n$ measure the departure of $\rm Ca_i^{peak}$, APD, and DI from those fixed point values during alternans. The cubic polynomial in $c_n$ in Eq.~(\ref{map1}) models a period-doubling bifurcation of the intracellular calcium dynamics with the amplitude of $\rm Ca_i^{peak}$ alternans increasing with the degree of calcium instability $r$. The term $~d_n$ in Eq. (\ref{map2}) incorporates APD-restitution. 
The other cross terms in Eqs. (\ref{map1})  and (\ref{map2}) model the bi-directional $\rm V_m- \rm Ca_i$ coupling taken to be positive in both directions ($\alpha>0$ and $\gamma>0$), corresponding to the typical case of locally in-phase APD and $\rm Ca_i^{peak}$ alternans.

\begin{figure}[t]
\centering
\addtolength{\belowcaptionskip}{-6mm}
\subfigure{
\epsfig{file =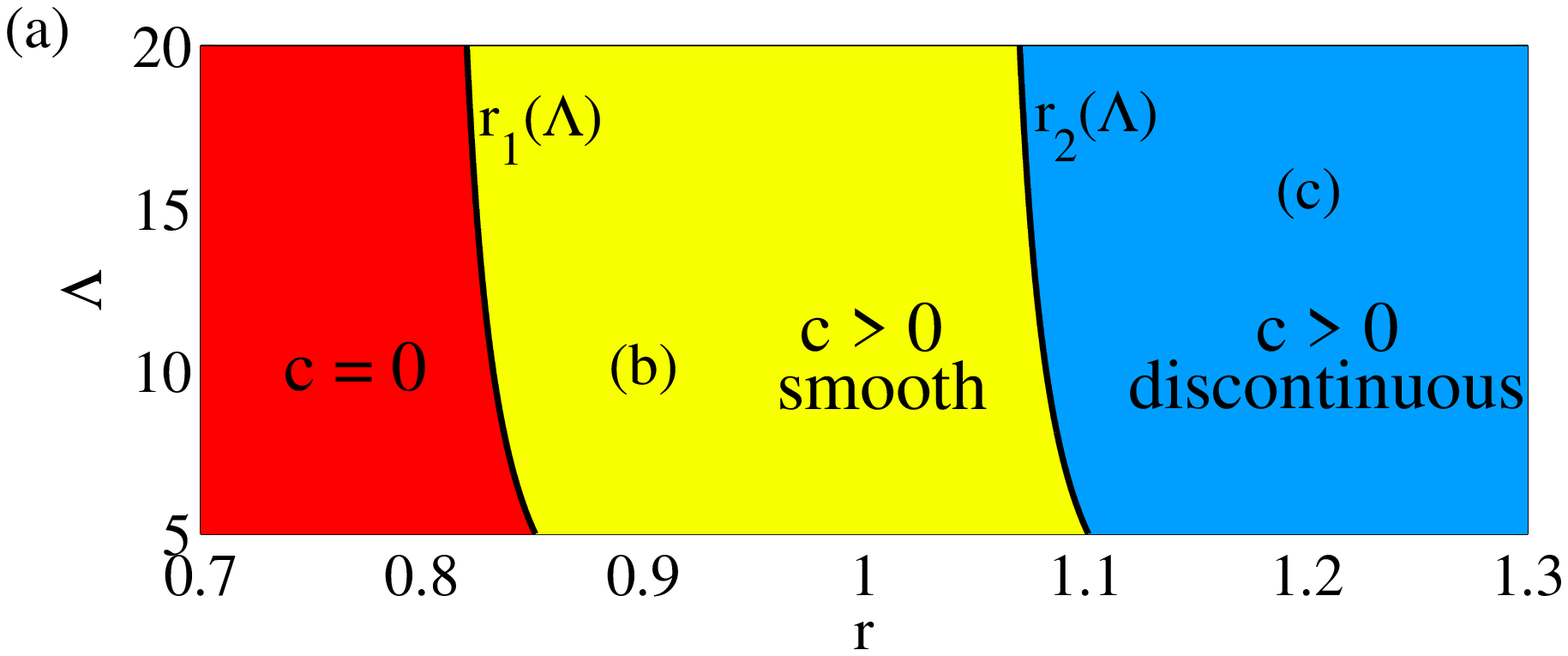, clip =,width=1.0\linewidth }
}
\subfigure{
\epsfig{file =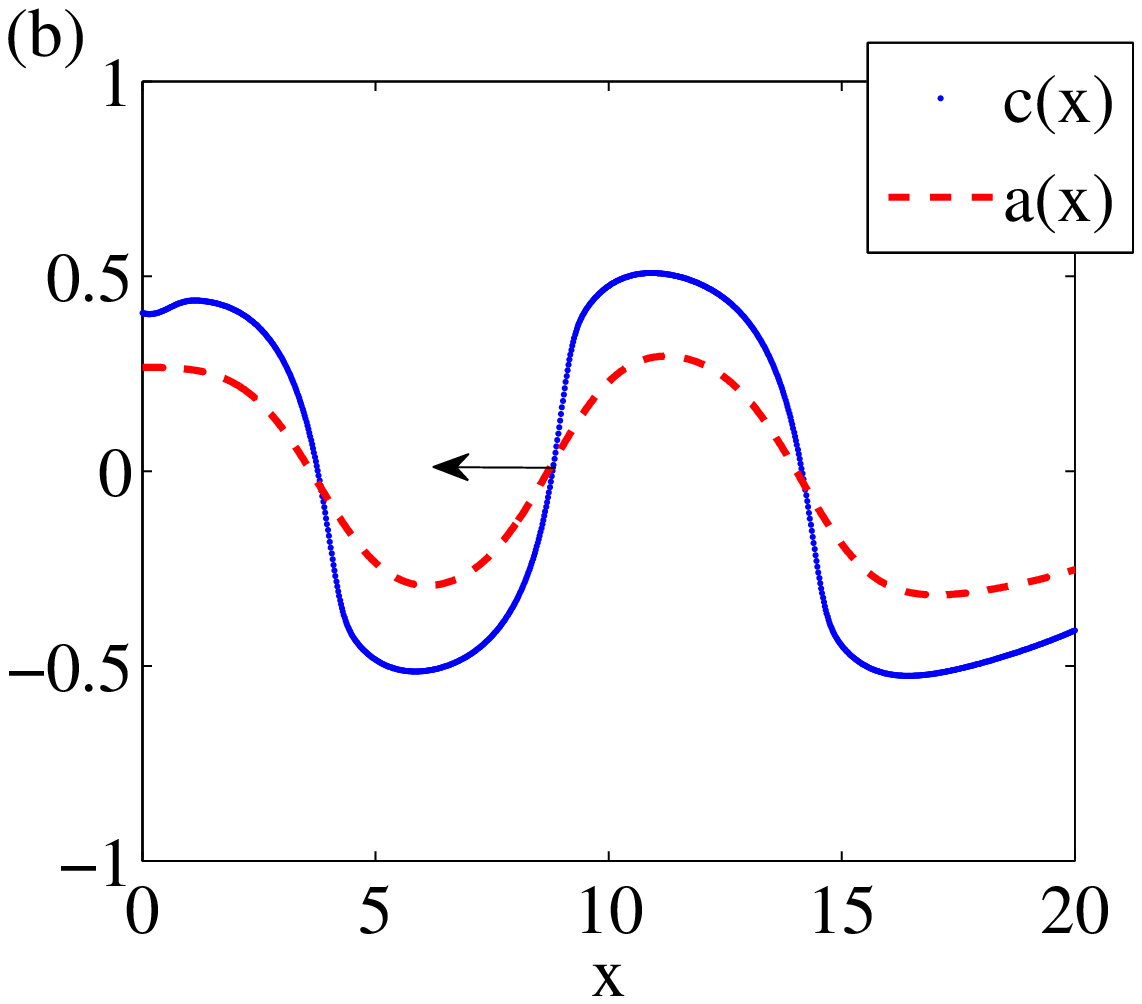, clip =,width=0.48\linewidth } \epsfig{file =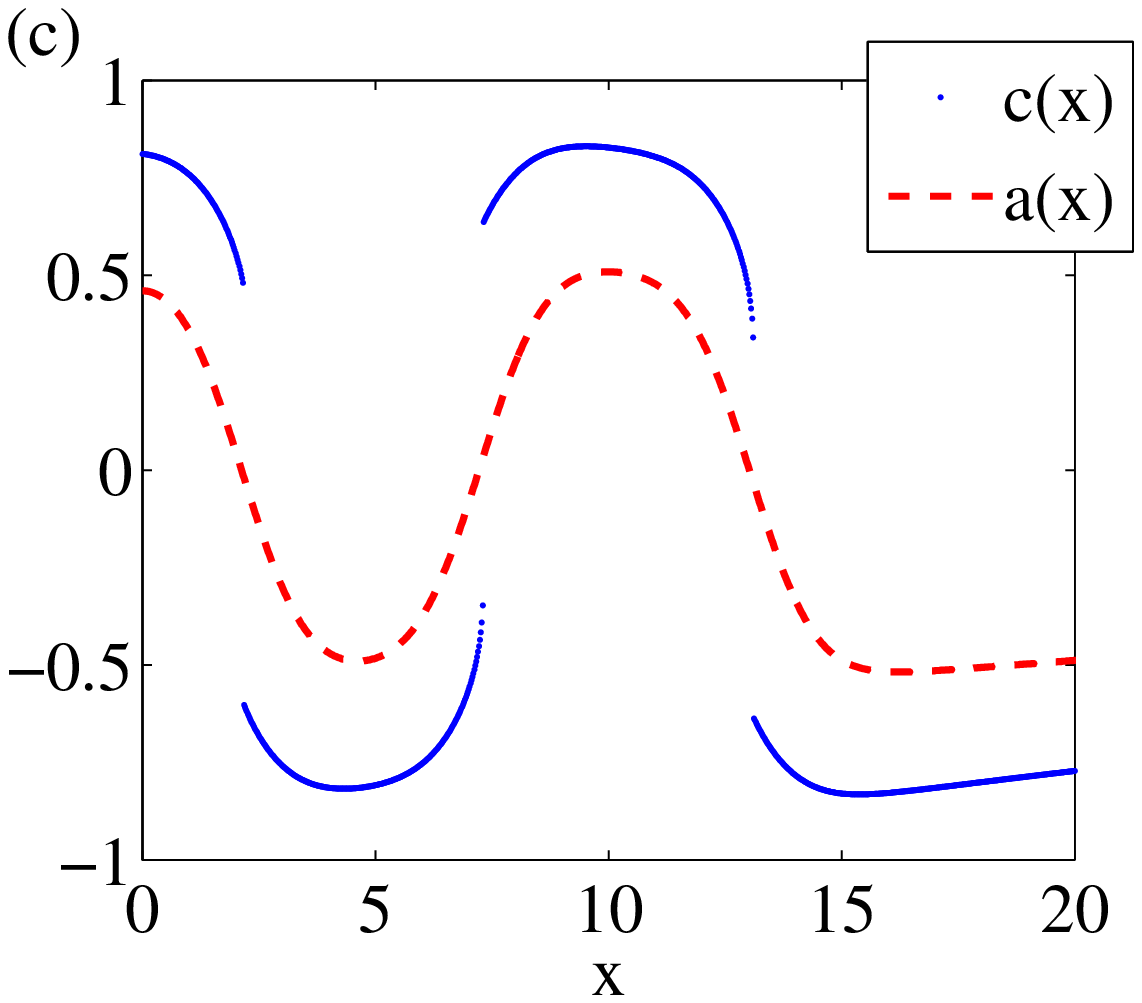, clip =,width=0.48\linewidth }
}
\subfigure{
\epsfig{file =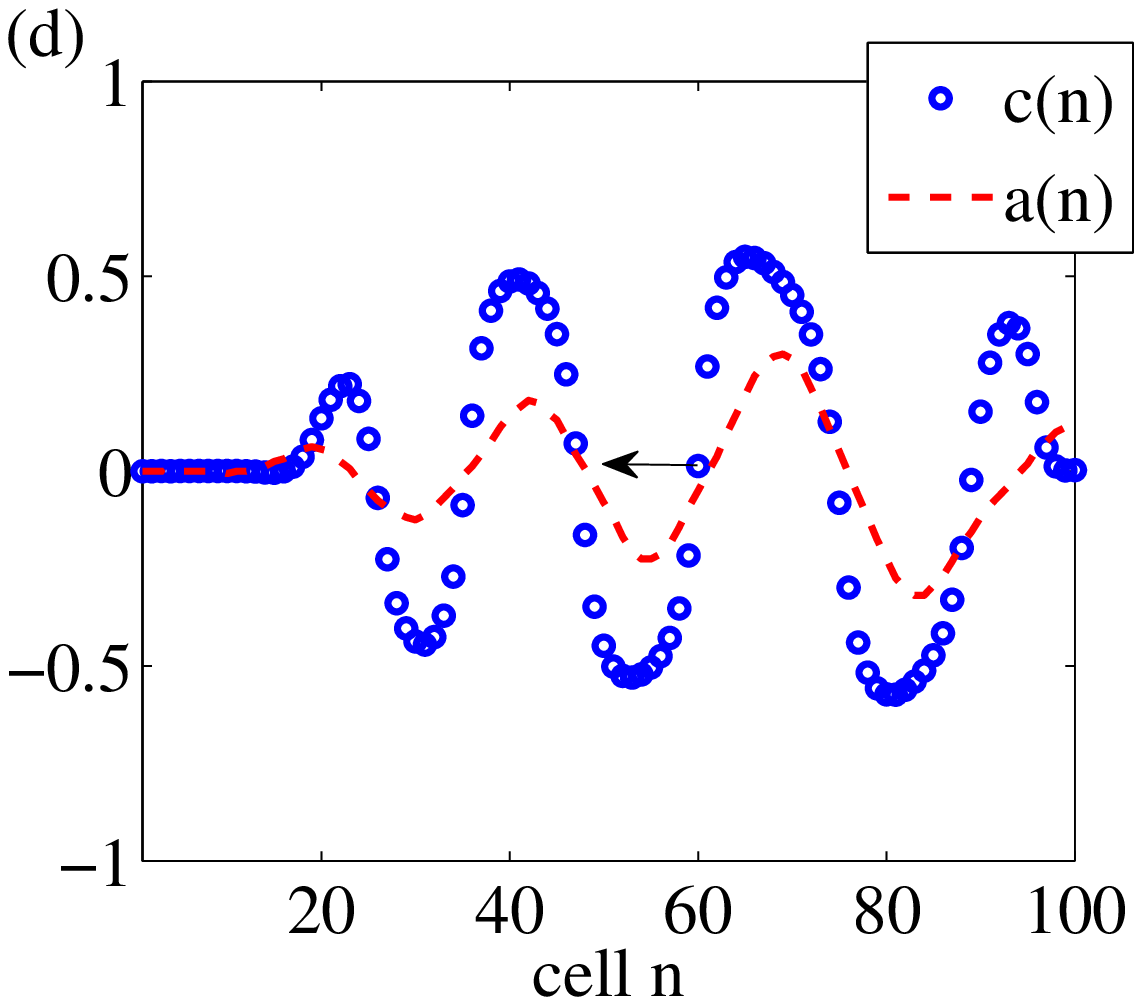, clip =,width=0.48\linewidth } \epsfig{file =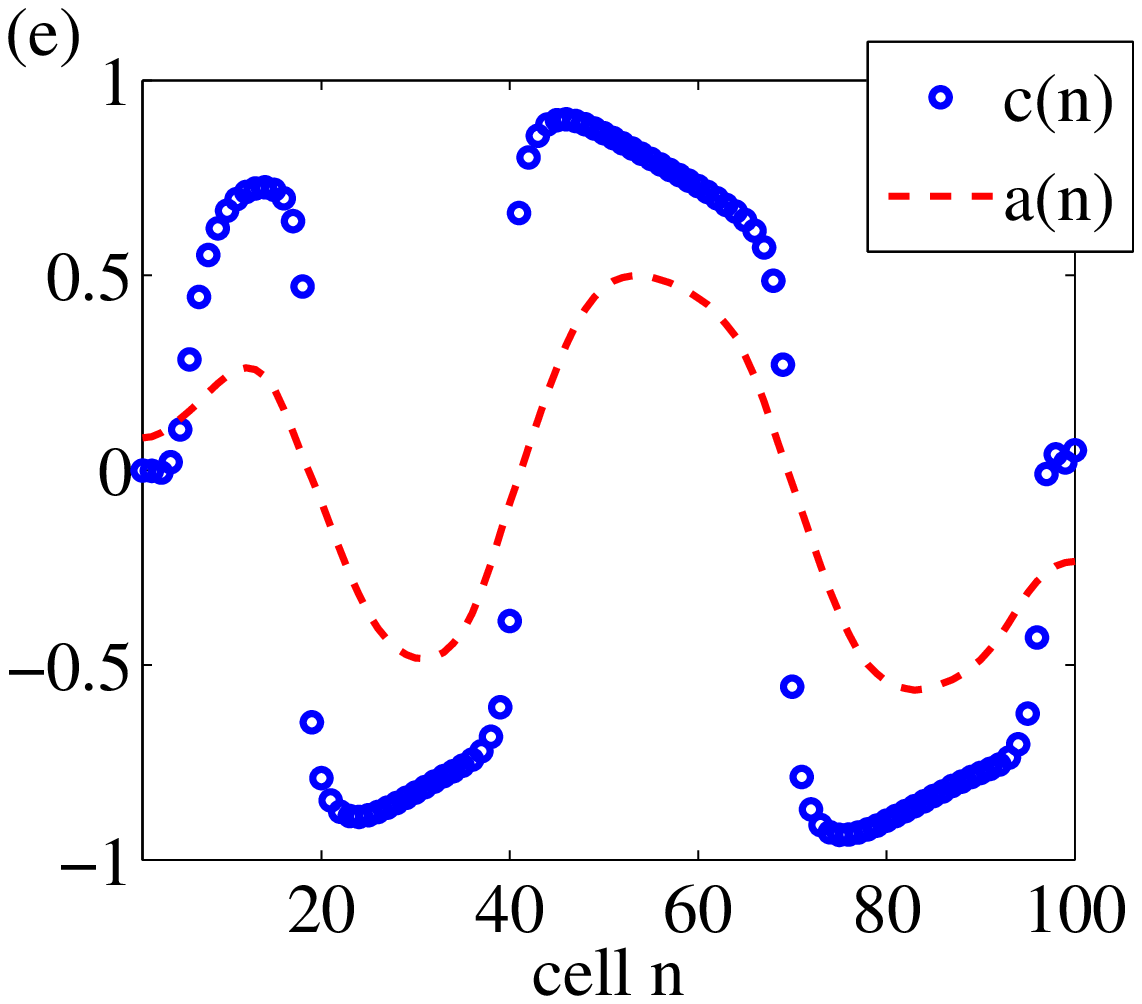, clip =,width=0.48\linewidth }
}
\caption{(Color online) (a) Nature of steady-state solutions in the $r$-$\Lambda$ plane. From left to right, no alternans ($c=0$), smooth calcium profiles ($c>0$, smooth), and discontinuous calcium profiles ($c>0$, discontinuous). (b) Smooth traveling and (c) discontinuous stationary $c(x)$ (blue) and $a(x)$ (dashed red) profiles from simulating Eqs.~(\ref{CV2})-(\ref{eqacv}), using $(r,\Lambda)=(0.9,10)$ and $(1.2,15)$, respectively. Alternans profiles obtained from a detailed ionic model \cite{Mahajan1} analogously showing (d) smooth traveling and (e) discontinuous stationary solutions.} \label{examples}
\end{figure}

To complete the derivation of maps describing the dynamics of $a_n$, $d_n$, and $c_n$, we linearize Eq. (\ref{period}) about the fixed point, which yields 
\begin{equation}
a_n+d_n=-\int_0^x\frac{dx'}{\Lambda}(d_n-d_{n-1})/2 \label{CV}
\end{equation}
where $\Lambda\equiv cv^{*2}/(2cv'^*)$. This linearization is formally valid as long as the amplitude of DI alternans induced by $\rm Ca_i$ alternans is small enough that we can locally neglect the curvature of the CV-restitution curve, or $|(r-1)^{1/2}\gamma A^*cv''^*/cv'^*|\ll 1$. Furthermore, we assume that the evolution of alternans amplitude is sufficiently slow that we can make the approximation $d_{n-1}\approx -d_n$. This assumption is valid close to the alternans bifurcation. 
Substituting $d_{n-1}= -d_n$ in Eq. (\ref{CV}) and differentiating both sides, we obtain a differential equation for $d_n(x)$ that can be solved exactly. Because the pacing rate is fixed at $x=0$, giving $a_n(0)+d_n(0)=0$, this yields
\begin{equation}
d_n(x)=-a_n(x)+e^{-x/\Lambda}\int_0^x \frac{dx'}{\Lambda}e^{x'/\Lambda}a_n(x').\label{CV2}
\end{equation}
The dynamics is completely specified by Eq. (\ref{CV2}) together with the maps obtained by inserting Eqs.~(\ref{map1})-(\ref{map2}) into Eqs.~(\ref{gmap1})-(\ref{gmap2}):
\begin{align}
 c_{n+1}&(x)=-rc_n(x)+c_n^3(x)+\alpha d_n(x),\label{eqccv}\\
 a_{n+1}&(x)=\int_0^L G(x,x')\left[\beta d_n(x')+\gamma c_{n+1}(x')\right]\ dx'.\label{eqacv}
\end{align}
Note that the pacing rate $\tau$ no longer appears in the final equations but is still contained implicitly in the fact that $D^*$, and hence the CV-restitution slope and $\Lambda$, can depend on $\tau$.
Also, since $d_{n-1}=-d_n$ in steady-state, Eqs.~(\ref{CV2})-(\ref{eqacv}) remain valid in steady-state even further from the bifurcation.

In Fig.~\ref{examples}, we present the results of different alternans behavior obtained from a numerical survey of Eqs. (\ref{CV2})-(\ref{eqacv}) where we vary systematically CV-restitution, which becomes shallower with increasing $\Lambda$, and the strength of $\rm Ca_i$-driven instability that increases with $r$. In Fig.~\ref{examples}(a) we summarize the nature of steady-state solutions in this parameter space. Small $r$ values yield no alternans solutions ($c=0$) where both steady-state $c(x)$ and $a(x)$ are identically zero. When $r$ is increased we find a first bifurcation at a value $r_1(\Lambda)$ where steady-state solutions for both $c(x)$ and $a(x)$ become non-zero  and form smooth waves ($c>0$, smooth). If the asymmetry of $G$, given by $w$, is not too small these waves are stationary, otherwise they move towards the pacing site with a constant velocity, as in the voltage dominated case \cite{EK}. For all work presented here $w=0$ was used, yielding traveling waves in the smooth regime. We found qualitatively similar results for positive $w$. 

When $r$ is increased further we find a second bifurcation at a value $r_2(\Lambda)$ where calcium alternans profiles become stationary and discontinuous at the nodes separating out of phase regions ($c>0$, discontinuous) while $a(x)$ remains smooth due to the smoothing effect of voltage diffusion. Example profiles of steady-state $c(x)$ (blue dots) and $a(x)$ (dashed red) are shown in Figs.~\ref{examples}(b) and (c) from the smooth and discontinuous regions, using $(r,\Lambda)=(0.9,10)$ and $(1.2,15)$, respectively. For all figures presented in this Letter we use parameters $\alpha=\gamma=\sqrt{0.4}$, $\beta=0$, $\xi=1$, and $w=0$. For comparison, in Figs.~\ref{examples}(d) and (e) we show $c(n)$ and $a(n)$ profiles inferred from numerical simulation of the detailed ionic model in Ref.~\cite{Mahajan1}, where $n$ indexes individual cells, using parameter values that give smooth traveling profiles and discontinuous stationary profiles, respectively. Traveling profiles have arrows indicating movement.

The onset of alternans at $r_1(\Lambda)$ is mediated by an absolute instability analogous to that studied in Ref.~\cite{EK} for the voltage-driven case. For $\beta=0$ a linear stability analysis yields thresholds of $r_1(\Lambda)=1-\eta+3\eta\xi^{2/3}/(4\Lambda^{2/3})$  and $1-\eta+\xi^2(w \Lambda)^{-1}$ for the instability of the traveling and stationary modes, respectively, where $\eta=\alpha\gamma$. Furthermore, the wavelength at onset is $4\pi\xi^{2/3}\Lambda^{1/3}/\sqrt{3}$ and $2\pi(w\Lambda)^{1/2}$ in the traveling and stationary cases, respectively, which agrees with the voltage-driven case in Ref.~\cite{EK}. Similar expressions can be obtained for $\beta\ne0$. Numerical simulations (not shown) are in good agreement with these theoretical results. 

We now concentrate on the discontinuous regime that is the primary focus of this letter. To characterize calcium alternans profiles in this regime [cf. Fig.~\ref{examples}(c)], we examine first stationary steady-state period-two profiles and substitute $c(x)=c_n(x)=-c_{n+1}(x)$ into Eq.~(\ref{eqccv}). After differentiating Eq.~(\ref{eqccv}) with respect to $x$ and some manipulations, we obtain
\begin{equation}\label{eqcprime}
 \Lambda c'(x)=\frac{c^3(x)-(r-1)c(x)-\alpha\Lambda a'(x)}{(r-1)-3c^2(x)}.
\end{equation}
Thus, when alternans grow from $c\sim0$ with $r>r_2(\Lambda)$, if $c(x)=c_-=\pm\sqrt{(r-1)/3}$ the derivative diverges and $c(x)$ becomes discontinuous. Through the discontinuity, the quantity $\alpha d(x)$ in Eq.~(\ref{eqccv}) remains smooth, so finding the other root of the cubic $(r-1)c=c^3+\alpha d$ gives the value of $c(x)$ at the latter end of the discontinuity. This gives $c(x)=c_+=\mp2\sqrt{(r-1)/3}$ and a total jump of amplitude $|c_+-c_-|=\sqrt{3(r-1)}$. To measure the asymmetry at a node, we introduce the quantity $\Delta\equiv ||c_+|-|c_-||/\sqrt{(r-1)/3}$. We will refer to a discontinuity where $c(x)$ jumps from $c_-=\pm\sqrt{(r-1)/3}$ to $c_+=\mp2\sqrt{(r-1)/3}$ as a {\it normal jump}. A remarkable property of this jump is that the limiting values $c_+$ and $c_-$ on either side of the node depend only on the strength $r$ of ${\rm Ca}_i$-driven instability, and is independent of all the other parameters $\Lambda$, $\eta$, $\beta$, $\xi$, and $w$. Experimentally, $r=1$ is the point in parameter space where an isolated myocyte paced with a periodic AP-clamp waveform, or a tissue paced at one point with negligible CV-restitution ($\Lambda=\infty$) bifurcates to alternans. Hence, the ratio $c_+/c_-$ can be used to deduce $r$ in tissue experiments or simulations under a finite effect of CV-restitution, and hence to relate single-cell and tissue behavior.

\begin{figure}[t]
\centering
\addtolength{\belowcaptionskip}{-5mm} 
\subfigure{
\epsfig{file =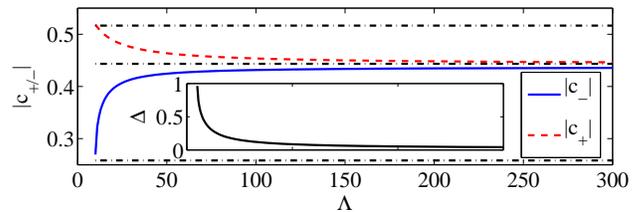, clip =,width=1.0\linewidth }
}
\caption{(Color online) Using $r=1.2$, $|c_-|$ (solid blue) and $|c_+|$ (dashed red) versus $\Lambda$ from an initial profile with $\Lambda=10$. Inset: $\Delta$.}\label{Lambda}
\end{figure}

When starting from the unstable base solution without alternans ($a=c=0$) in the regime $r>r_2(\Lambda)$, SDA forms dynamically as a periodic pattern of discontinuous nodes with normal jumps. A  unique  feature of SDA evolution in this regime, which is entirely absent for $V_m$-dominated instability, is that both the node positions and alternans profiles can depend strongly on the {\it history} of how the parameters $r$ and $\Lambda$ are varied. If $\Lambda$ or $r$ are increased starting from a profile with normal jumps, the position of the nodes remains constant, but the shape of the profile deforms in such a way that the jump in ${\rm Ca}_i$-alternans profile becomes symmetrical about the node, i.e.  both $|c_-|$ and $|c_+|$ approach the same limiting value $|c_{\pm}| =\sqrt{r-1}$ where $\Delta$ vanishes. This shows that, if initial conditions contain discontinuous nodes, jumps need not be normal in steady-state if $c(x)=c_-=\pm\sqrt{(r-1)/3}$ is not attained. This is shown in Fig.~\ref{Lambda} where we plot $|c_-|$ (solid blue) and $|c_+|$ (dashed red) using $r=1.2$ as we increase $\Lambda$ from $10$. As $\Lambda$ is increased, $|c_-|$ and $|c_+|$ tend toward one another. We superimpose theoretical values of $|c_\pm|$ for the normal jump and $\Lambda^{-1}=0$ cases in dot-dashed black for comparison, noting that $|c_-|,|c_+|$ vary smoothly between these values for intermediate values of $\Lambda$. In the inset we plot $\Delta$ versus $\Lambda$, noting that $\Delta\to 0$ as $\Lambda\to\infty$. If $\Lambda$ is then decreased back until it reaches its original value (not shown), the profile recovers its original shape. However, if $\Lambda$ or $r$ are decreased starting at a point where the jumps are normal, the pattern close to the node preserves its shape, but the node moves towards the pacing site. Importantly, if $\Lambda$ or $r$ are increased back after the node has moved, the node does not return to its original position, but rather its shape will deform to become symmetrical as described above. Since no parameter change can induce the node to move away from the pacing site, node motion is {\it unidirectionally pinned}. We note that we have also observed unidirectional pinning in our ionic model simulations~\cite{Mahajan1}.

When the node is unpinned, we find that the location of the first node, denoted $x_1$, scales linearly with $\Lambda$, suggesting that the node spacing is independent of electrotonic coupling. This linear scaling with $\Lambda$ in the discontinuous regime is to be contrasted with the scaling of the node spacing for smooth alternans profiles  (e.g. $\xi^{2/3}\Lambda^{1/3}$ for $w = 0$), which depends strongly on electrotonic coupling.  Physically, this linear scaling reflects the fact that electrotonic coupling has a negligible effect on the outer scale where the alternans profile varies slowly on a scale $\sim \Lambda$, and only becomes relevant on a scale $\sim \xi$ near the nodes. This only adds a subdominant correction of order $\xi$ to the $x_1\sim \Lambda$ scaling. Mathematically, it can be related to the fact that $\Lambda$ scales out of Eq.~(\ref{eqcprime}) in the limit $\Lambda \gg \xi$ if one uses the scaled variable $\tilde{x}=x/\Lambda$ instead of $x$.

\begin{figure}[t]
\centering
\addtolength{\belowcaptionskip}{-6mm}
\subfigure{
\epsfig{file =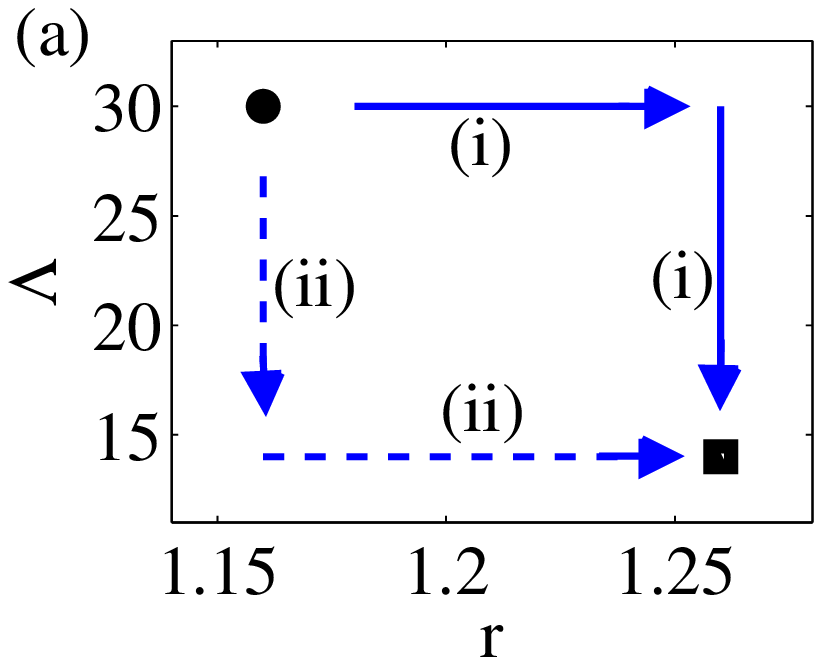, clip =,width=0.48\linewidth } \epsfig{file =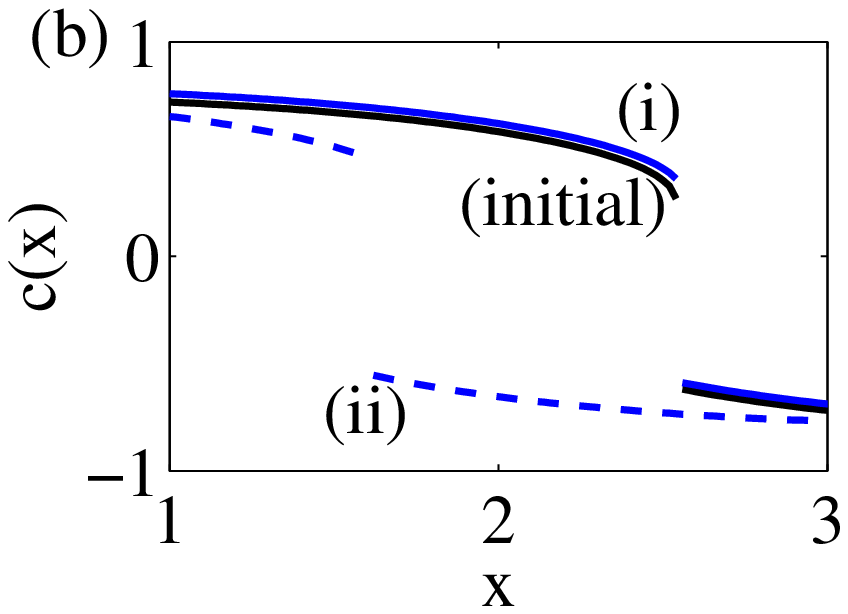, clip =,width=0.48\linewidth }
}
\subfigure{
\epsfig{file =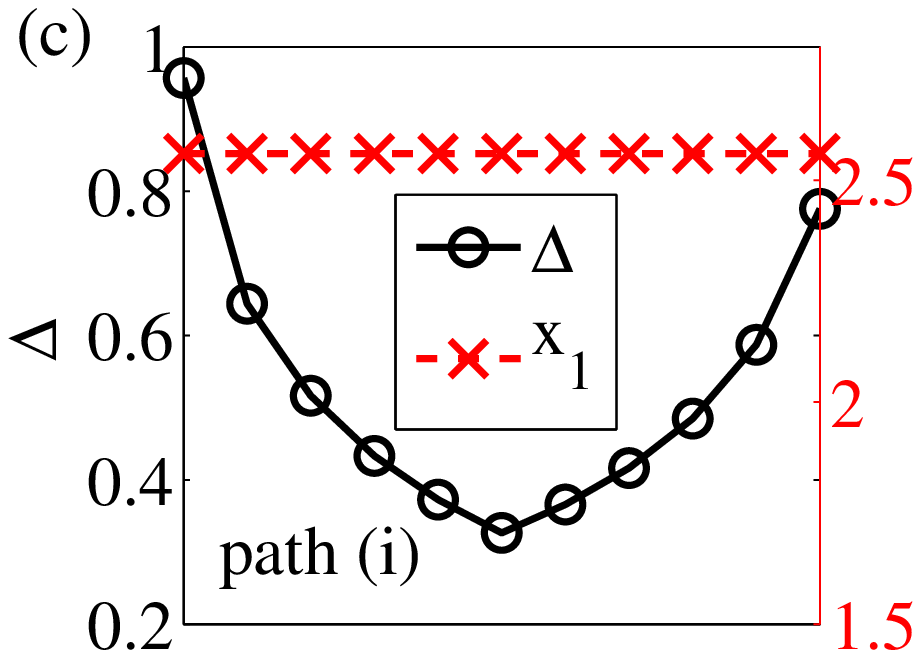, clip =,width=0.48\linewidth } \epsfig{file =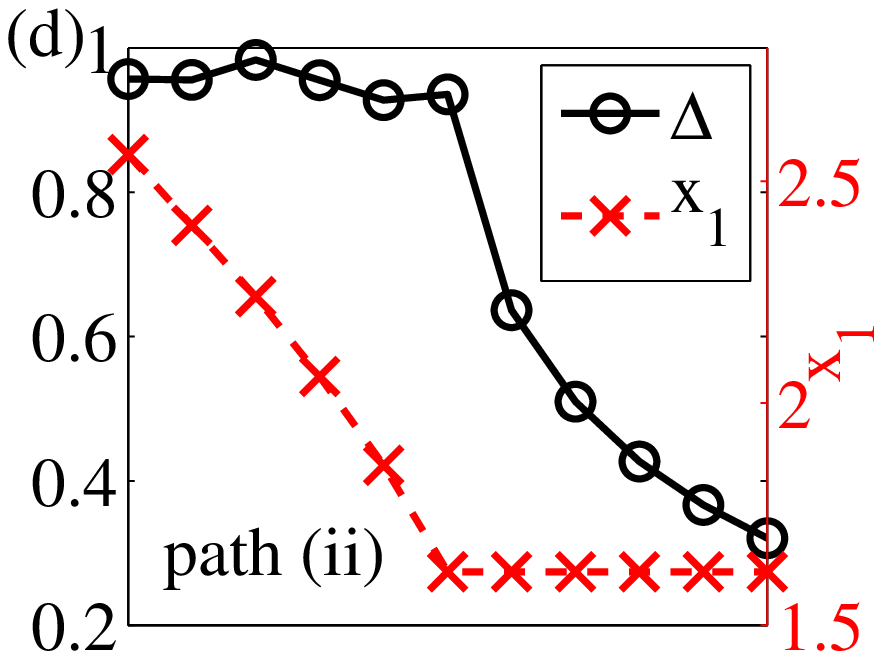, clip =,width=0.48\linewidth }
}
\caption{(Color online) (a) Paths (i) and (ii) (solid and dashed) in $(r,\Lambda)$. (b) Initial profile (solid black) and final profiles (solid and dashed blue) after moving along path (i) and (ii). (c), (d) Asymmetry $\Delta$ (circles) and $x_1$ (crosses) along paths (i) and (ii), respectively.}\label{rLambda}
\end{figure}

To investigate the consequences of unidirectional pinning, we investigated the pattern evolution in response to multiple parameter changes. Namely, we changed $r$ and $\Lambda$ following two different paths that connect the same points in $(r,\Lambda)$. Starting with a pattern with normal jumps at $(r,\Lambda)=(1.16,30)$, we move to $(1.26,14)$ first by increasing $r$ and then decreasing $\Lambda$ [path (i)] and vice versa [path (ii)]. Despite the same start and end parameters, the resulting profile characteristics vary significantly depending on the path followed, as shown in Fig.~\ref{rLambda}. In Fig.~\ref{rLambda}(a) paths (i) and (ii) are denoted by solid and dashed blue arrows, respectively, with the start and endpoints denoted as a black circle and square. In Fig. \ref{rLambda}(b) we zoom in on the first node of the initial profile (solid black curve) and final profiles after moving along path (i) (solid blue) and (ii) (dashed blue). Consistent with our summary above, $x_1$ remains constant along path (i) while $\Delta$ decreases as $r$ is increased, after which $\Delta$ increases as $\Lambda$ is decreased. Along path (ii) $x_1$ decreases with $\Lambda$, then remains constant while $\Delta$ decreases as $r$ is increased. This is shown in Fig.~\ref{rLambda} (c) and (d), which show $\Delta$ (black circles) and $x_1$ (red crosses) measured along paths (i) and (ii), respectively.

In conclusion, we have extended our basic theoretical understanding of SDA dynamics to the important case of calcium-driven instability. Furthermore we have made a number of new experimentally testable predictions, which we have validated by detailed ionic model simulations. The main prediction is that node motion becomes unidirectionally pinned when the ${\rm Ca_i}$ alternans profile becomes spatially discontinuous above a threshold of ${\rm Ca_i}$-driven instability. This prediction could be tested by first increasing progressively the pacing rate (decreasing the inverse restitution slope $\Lambda$), thereby causing the node to move towards the pacing site, as has already been observed in some experiments on APD SDA \cite{Hayashi,ZivMironov}, and then decreasing the pacing rate to its original value. If the ${\rm Ca}_i$-alternans profile exhibits a jump at the node, the node should remain stationary. Increasing pacing rate can also cause several parameters to change, including the degree of ${\rm Ca_i}$-driven instability. However, we have shown that history-dependent SDA evolution is robust to multiple parameter changes  (Fig. \ref{rLambda}), and hence should be observable in more complex situations. We emphasize that unidirectional pinning is a purely dynamical phenomenon independent of intrinsic tissue heterogeneities, which can also potentially pin node motion. However, we expect pinning due to tissue heterogeneities to be generally bi-directional, and hence distinguishable from unidirectional dynamically-induced pining.  A second prediction is that the spatial jump in ${\rm Ca_i}$-alternans amplitude displays remarkably universal features. The magnitude and asymmetry of this jump are insensitive to most parameters except the degree of ${\rm Ca_i}$-driven instability, and both quantities are generally history-dependent.

Unidirectional pinning generally makes it harder to eliminate SDA by node motion once they are formed. We therefore expect SDA to be more arrhythmogenic for ${\rm Ca_i}$- than $V_m$-dominated instability. Given that alternans are believed to be predominantly ${\rm Ca_i}$-driven in common pathologies such as heart failure, SDA may play an even more important role than previously thought in such pathologies.

The work of A.K. and J.G.R. was supported in part by National Institutes of Health grant No. P01 HL078931.

\bibliographystyle{plain}

%\bibliography{AltRef}

%footnote For convenience of normalization, $\xi$ is defined to be $\sqrt{2}$ larger than in Ref. (EchebarriaKarma2002\&2007).)

%Pastore JM, Girouard SD, Laurita KR, Akar FG, Rosenbaum DS.
%Mechanism linking T-wave alternans to the genesis of cardiac fibrillation.
%Circulation. 1999;99:1385Ð1394.

%ÒOrigin of complex behaviour of spatially discordant alternans in a transgenic rabbit model of type 2 long QT syndromeÓ, O. Ziv, E. Morales, Y.-K. Song, X. Peng, K.E. Odening, A.E. Buxton, A. Karma, G. Koren, and B.-R. Choi, J Physiol   587, 4661-4680 (2009).
 
%\bibitem{chemical}	J. S. Park, and K. J. Lee, Physical Review Letters {\bf 83}, 5393 (1999); J. S. Park, Sung-Jae Woo, and K. J. Lee, Phys. Rev. Lett. {\bf 93}, 098302 (2004).

%\bibitem{granular} Francisco Melo, Paul B. Umbanhowar, and Harry L. Swinney. Hexagons,
%kinks, and disorder in oscillated granular layers. Physical Review Letters,
%75(21):3838-3842, November 1995. 

%Qu Shiferaw Weiss PHYSICAL REVIEW E 75, 011927 2007

%M.R. Guevara, G. Ward, A. Shrier, L. Glass. Electrical alternans and period-doubling bifurcations. IEEE Computers in Cardiology, 167-170 (1984).

%J. N. Weiss et al., Circ. Res. {\bf 98}, 1244 (2006).

 %A. Karma, and R. F. Gilmour, Physics Today {\bf 60}, 51 (2007) 
 
%M. Courtemanche, L. Glass, J.P. Keener. Instabilities of a propagating pulse in a ring of excitable media. Physical Review Letters 70, 2182-2185 (1993)

\end{document}